\documentclass[bibyear]{aa} 
\begin{document} 

\title{Blobs, spiral arms, and a possible planet around HD~169142
   \thanks{Based on data collected at the European Southern Observatory,
Chile (ESO Program 1100.C-0481)}}

\author{R. Gratton
          \inst{1},
          R. Ligi
          \inst{2},
          E. Sissa
          \inst{1},
          S. Desidera
          \inst{1},
          D. Mesa
          \inst{1},
          M. Bonnefoy
          \inst{3},
          G. Chauvin
          \inst{3,4},
          A. Cheetham
          \inst{5},
          M. Feldt
          \inst{6},
          A.M. Lagrange
          \inst{3},
          M. Langlois
          \inst{7,8},
          M. Meyer
          \inst{9,10},
          A. Vigan
          \inst{7},
          A. Boccaletti
          \inst{11},
          M. Janson
          \inst{6,12},
          C. Lazzoni
          \inst{1}
          A. Zurlo
          \inst{7,13},
          J. De Boer
          \inst{14},
          T. Henning
          \inst{6},
          V. D'Orazi
          \inst{1},
           L. Gluck
          \inst{3},
           F. Madec
          \inst{7},
           M. Jaquet
          \inst{7},
           P. Baudoz
          \inst{11},
           D. Fantinel     
          \inst{1},
           A. Pavlov
          \inst{6},
          \and
           F. Wildi          
          \inst{5}}
   

\institute{$^{1}$Osservatorio Astronomico di Padova - INAF \email{raffaele.gratton@inaf.it}\\
$^{2}$Osservatorio Astronomico di Brera - INAF               \email{roxanne.ligi@inaf.it}\\
$^{3}$Univ. Grenoble Alpes, CNRS, IPAG, F-38000 Grenoble, France\\ 
$^{4}$Unidad Mixta Internacional Franco-Chilena de Astronom\'ia,CNRS/INSU UMI 3386 and Departamento de Astronom\'ia, Universidad de Chile, Casilla 36-D, Santiago, Chile\\
$^{5}$Geneva Observatory, University of Geneva, Chemin des Mailettes 51, 1290 Versoix, Switzerland\\
$^{6}$Max Planck Institute for Astronomy, Konigstuhl 17, D-69117 Heidelberg, Germany\\
$^{7}$Aix Marseille Universit\'e, CNRS, LAM (Laboratoire d'Astrophysique de Marseille) UMR 7326, 13388 Marseille, France\\
$^{8}$CRAL, UMR 5574, CNRS, Universit de Lyon, Ecole Normale Superieure de Lyon, 46 Alle d'Italie, F-69364 Lyon Cedex 07, France\\
$^{9}$Institute for Astronomy, ETH Zurich, Wolfgang-Pauli-Strasse 27, 8093 Zurich, Switzerland\\
$^{10}$Department of Astronomy, University of Michigan, Ann Arbor, MI 48109, US\\
$^{11}$LESIA, Observatoire de Paris, PSL Research University, CNRS, Sorbonne Universite, UPMC Univ. Paris 06, Univ. Paris Diderot, Sorbonne Paris Cit, 5 place Jules Janssen, 92195 Meudon, France\\
$^{12}$Department of Astronomy, Stockholm University, SE-106 91 Stockholm, Sweden\\
$^{13}$N\'ucleo de Astronoma, Facultad de Ingenier\'ia, Universidad Diego Portales, Av. Ejercito 441, Santiago, Chile\\
$^{14}$Leiden Observatory, Universiteit Leiden}

   \date{Received ; accepted }

 
  \abstract
   {Young planets are expected to cause cavities, spirals, and kinematic perturbations in protostellar disks that may be used to infer their presence. However, a clear detection of still-forming planets embedded within gas-rich disks is still rare.}
   {HD~169142 is a very young Herbig Ae-Be star surrounded by a pre-transitional disk, composed of at least three rings. While claims of sub-stellar objects around this star have been made previously, follow-up studies remain inconclusive. The complex structure of this disk is not yet well understood.}
   {We used the high contrast imager SPHERE at ESO Very large Telescope to obtain a sequence of high-resolution, high-contrast images of the immediate surroundings of this star over about three years in the wavelength range 0.95-2.25~$\mu$m. This enables a photometric and astrometric analysis of the structures in the disk.}
   {While we were unable to definitively confirm the previous claims of a massive sub-stellar object at 0.1-0.15 arcsec from the star, we found both spirals and blobs within the disk. The spiral pattern may be explained as due to the presence of a primary, a secondary, and a tertiary arm excited by a planet of a few Jupiter masses lying along the primary arm, likely in the cavities between the rings. The blobs orbit the star consistently with Keplerian motion, allowing a dynamical determination of the mass of the star. While most of these blobs are located within the rings, we found that one of them lies in the cavity between the rings, along the primary arm of the spiral design.}
   {This blob might be due to a planet that might also be responsible for the spiral pattern observed within the rings and for the cavity between the two rings. The planet itself is not detected at short wavelengths, where we only see a dust cloud illuminated by stellar light, but the planetary photosphere might be responsible for the emission observed in the K1 and K2 bands. The mass of this putative planet may be constrained using photometric and dynamical arguments. While uncertainties are large, the mass should be between 1 and 4 Jupiter masses. The brightest blobs are found at the 1:2 resonance with this putative planet.  }


\authorrunning{R. Gratton et al.}
\titlerunning{Blobs around HD~169142}
\maketitle
%

\section{Introduction}

Planet formation occurs in disks around young stellar objects. Interactions between planets and disks are very complex. Young planets are expected to cause rings, cavities, spirals, and disturbances in the velocity field and other features in the disk, which in turn may be used to infer the presence of these young planets. In the past few years, much evidence about this phase of planet formation has been accumulated because high-resolution images in the millimeter and sub-millimeter wavelength ranges have been provided by the Very Large Array (VLA) and the Atacama Large Millimeter Array (ALMA) (see e.g. the case of HL Tau: \cite{ALMA2015}), and by high-contrast imagers such as the Gemini Planet Imager (GPI: \cite{Macintosh2014}) and SPHERE (Spectro-
Polarimetic High contrast imager for Exoplanets REsearch: (\cite{Beuzit2008}; see, e.g., \cite{Avenhaus2018}). The literature on indirect evidence of the presence of planets is now becoming very rich, and nearby young stars surrounded by gas-rich disks are intensively studied for this purpose. In most cases, available data cannot fully eliminate alternative hypotheses, or the data have ambiguous interpretations (see, e.g., \cite{Bae2018} and \cite{Dong2018}), although strong indirect evidence of the presence of planets from local disturbances of the velocity field have recently been considered for the case of HD~163296 (\cite{Pinte2018, Teague2018}). In general, small grains are thought to be more strongly coupled with gas and are thus less sensitive to radial drift and concentration that can strongly affect large grains (see the discussion in \cite{Dipierro2018}). For this reason, observations at short wavelengths provide an important complementary view of what can be seen with ALMA. On the other hand, a direct detection of still-forming planets embedded within primordial gas-rich disks, which is expected to be possible with high-contrast imaging in the near infrared (NIR), is still scarce; remarkable cases are LkCa-15 (\cite{Kraus2012, Sallum2015}) and PDS-70 (\cite{Keppler2018, Muller2018, Wagner2018}). In particular, in this second case, a clear detection of an accreting planet in the cavity between the inner and outer ring was obtained, making it an archetype for planet formation and planet-disk interactions. However, many cases remain ambiguous; a classical example is HD~100546 (see, e.g., \cite{Quanz2013a, Currie2014, Quanz2015, Currie2015, Rameau2017, Sissa2018}).

\begin{figure*}
\centering
\includegraphics[width=18.8cm]{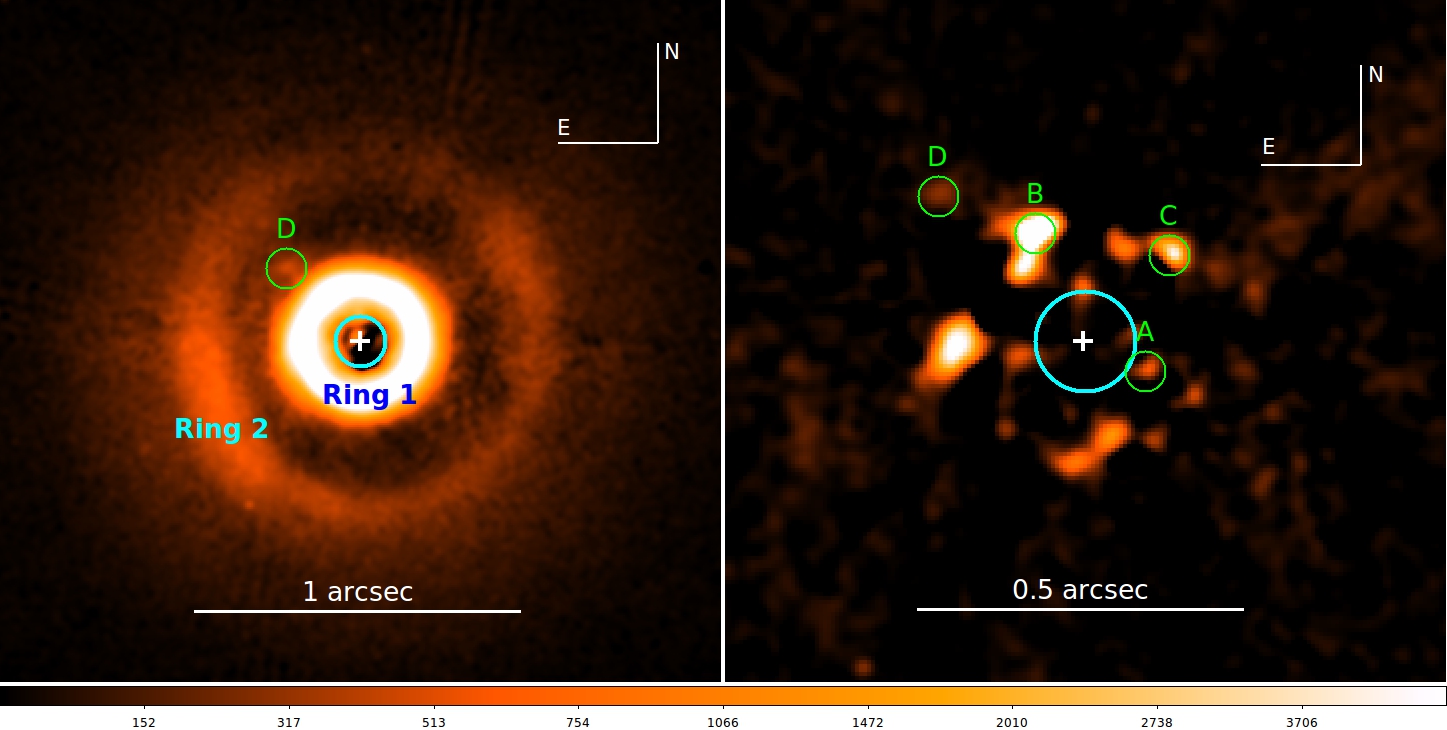}
\caption{View of the surroundings of HD~169142 obtained from polarimetric observations: the left panel shows the Q$_\Phi$\ image in the J band acquired with SPHERE (\cite{Pohl2017}) on a linear scale. The two rings are clearly visible. The right panel shows a pseudo-ADI image of the inner regions obtained by differentiating the Q$_\Phi$\ image (see \cite{Ligi2017}, for more details). The white cross marks the position of the star, and the cyan circle shows the size of the coronagraphic mask. The other labels refer to the blobs we discuss in this paper that are also visible in these images. The color scale of the differential image is five times less extended to show the faint structures better. In both panels, N is up and E to the left; a segment represents 1 and 0.5 arcsec in the left and right panel, respectively. }
\label{f:pdi}
\end{figure*}

HD~169142 is a very young Herbig Ae-Be star with a mass of 1.65-2~M$_\odot$ and an age of 5-11~Myr (\cite{Blondel2006, Manoj2007}) that is surrounded by a gas-rich disk ($i=13$~degree; \cite{Raman2006}; $PA=5$~degree; \cite{Fedele2017}) that is seen almost face-on. The parallax is $8.77\pm 0.06$~mas (GAIA  DR2 2018). Disk structures dominate the inner regions around HD~169142 (see, e.g., \cite{Ligi2017}). Figure~\ref{f:pdi} shows the view obtained from polarimetric observations: the left panel shows the Q$_\Phi$\ image in the J band obtained by \cite{Pohl2017} using SPHERE on a linear scale, and the two rings are clearly visible. The right panel shows a pseudo-ADI image of the inner regions obtained by differentiating the  Q$_\Phi$\ image (see \cite{Ligi2017}, for more details). \cite{Biller2014} and \cite{Reggiani2014} discussed the possible presence of a point source candidate at small separation ($<0.2$~arcsec from the star). However, the analysis by Ligi et al. based on SPHERE data does not support or refute these claims; in particular, they suggested that the candidate identified by Biller et al. might be a disk feature rather than a planet. Polarimetric images with the adaptive optics system NACO at the Very Large Telescope (VLT) (\cite{Quanz2013b}), SPHERE (\cite{Pohl2017, Bertrang2018}) and GPI (\cite{Monnier2017}) show a gap at around 36~au, with an outer ring at a separation $>$40~au from the star. This agrees very well with the position of the rings obtained from ALMA data (\cite{Fedele2017}); similar results were obtained from VLA data (\cite{Osorio2014, Macias2017}). We summarize this information about the disk structure in Table~\ref{t:rings} and call the ring at 0.17-0.28 arcsec from the star Ring 1 and the ring at 0.48-0.64 arcsec Ring 2. We remark that in addition to these two rings, both the spectral energy distribution (\cite{Wagner2015}) and interferometric observations (\cite{Lazareff2017, Chen2018}) show  an inner disk at a separation smaller than 3 au. This inner disk is unresolved from the star in high-contrast images and consistent with ongoing accretion from it onto the young central star. While the cavities between the rings seem devoid of small dust, some gas is present there (\cite{Osorio2014, Macias2017,  Fedele2017}). \cite{Fedele2017} and \cite{Bertrang2018} have suggested the possibility that the gap between Rings 1 and 2 is caused by a planet with a mass slightly higher than that of Jupiter. However, this planet has not yet been observed, possibly because it is at the limit of or beyond current capabilities of high-contrast imagers. On the other hand, \cite{Bertrang2018} found a radial gap in Ring 1 at PA$\sim 50$\ degree that might correspond to a similar radial gap found by \cite{Quanz2013b} at PA$\sim 80$\ degree. The authors noted that if this correspondence were real, then this gap might be caused by a planet at about 0.14 arcsec from the star. So far, this planet has not been unambiguously detected either.

\begin{table*}
\centering
\caption{Rings around HD169142 from the literature}
\begin{tabular}{lccc}
\hline
\hline
Instrument & Source & Ring 1 & Ring 2 \\
           &        & arcsec & arcsec \\
\hline
ALMA          & \cite{Fedele2017}   & 0.17-0.28 & 0.48-0.64 \\
VLA           & \cite{Osorio2014}   & 0.17-0.28 & 0.48-     \\
SUBARU-COMICS & \cite{Honda2012}    &   0.16-   &           \\
NACO          & \cite{Quanz2013b}   & 0.17-0.27 & 0.48-0.55 \\
SPHERE-ZIMPOL & \cite{Bertrang2018} & 0.18-0.25 & 0.47-0.63 \\
SPHERE-IRDIS  & \cite{Pohl2017}     & 0.14-0.22 & 0.48-0.64 \\
GPI           & \cite{Monnier2017}  &    0.18   &    0.51   \\
\hline
\end{tabular}
\label{t:rings}
\end{table*}

In this paper, we pursue a new view on the subject through analyzing high-contrast images. In particular, we underline that while polarimetric observations in the NIR and millimeter observations are best to reveal the overall structure of the disk, pupil-stabilized NIR observations where angular differential imaging can be applied may reveal fainter structures on a smaller scale. The risk of false alarms inherent to the image-processing procedures used in high-contrast imaging can be mitigated by comparing different sets of observations taken at intervals of months or years. In the case of HD~169142, this is exemplified by the study of \cite{Ligi2017}, who identified a number of blobs within Ring 1. We have now accumulated a quite consistent series of observations of this star with SPHERE that extends the set of data considered by Ligi et al. The observations have a comparable limiting contrast so that we may try to combine this whole data set to improve our knowledge of this system. The combination of different data sets acquired over a few years offers several advantages. In addition to verification of previous claims, we might try to detect persistent features around HD~169142 using a coincidence method to obtain a combined image that is deeper than the individual images and allows a quantitative discussion of the false-alarm probability of detected features. The expected orbital motion needs to be taken into account in this.

In Section 2 we describe observation and analysis methods. In Section 3 we present the main results about the blobs we detect around the star. in Section 4 we discuss the spiral arms within the disk and the possible connection to the blobs. Conclusions are given in Section 5.

\begin{figure*}
\centering
\includegraphics[width=18.8truecm]{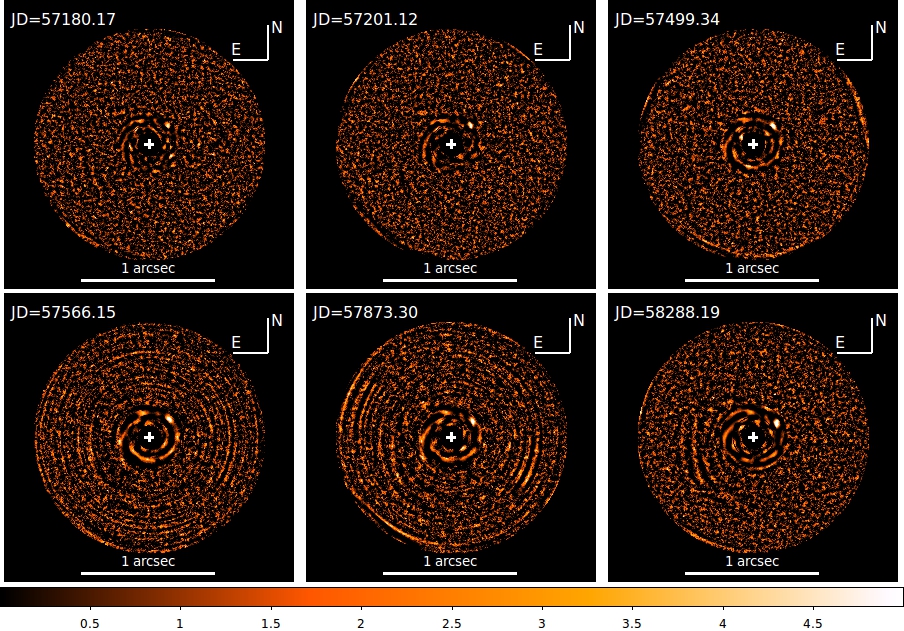}
\caption{Signal-to-noise ratio maps for the individual epochs for IFS. The individual S/N maps are obtained from the ASDI PCA algorithm using 50 components and making a median over the wavelength (see \cite{Mesa2015}). In the upper row we show in the left panel JD=57180.17, in the middle panel JD=57201.12, and in the right panel JD=57499.34.  In the lower row we show in the  left panel JD=57566.15, in the middle panel JD=57873.30, and in the right panel JD=58288.19. In all panels, the central 0.1 arcsec is masked, the solid white line at the bottom represents 1 arcsec, the white cross represents the position of the star. N is up and E to the left.}
\label{f:individual}
\end{figure*}

\section{Observation and data analysis}

Data were acquired with the SPHERE high-contrast imager (\cite{Beuzit2008}) at the ESO VLT Unit Telescope 3 within the guaranteed time observations used for the SHINE (SpHere INfrared survey for Exoplanets) survey (\cite{Chauvin2017}). Data acquired up to 2017 have been described in \cite{Ligi2017}. Here we add new data acquired in 2018 and study the system anew using different ways to combine different images. In these observations, we used SPHERE with both the Integral Field Spectrograph (IFS : \cite{Claudi2008}) and the Infra-Red Dual Imaging and Spectrograph (IRDIS: \cite{Dohlen2008, Vigan2010} simultaneously. IFS was used in two modes: Y-J, that is, with spectra from 0.95 to 1.35~$\mu$m and a resolution of $R\sim 50$; and Y-H, with spectra from 0.95 to 1.65~$\mu$m and a resolution of $R\sim 30$. When IFS was in Y-J mode, IRDIS observed in the H2-H3 narrow bands (1.59 and 1.66~$\mu$m, respectively); when IFS was in Y-H mode, IRDIS observed in K1-K2 bands (2.09 and 2.25~$\mu$m, respectively). Hereafter we mainly consider data acquired with the IFS;  IRDIS data are considered for the photometry in the K1-K2 bands. We considered the six best observations obtained for HD169142 (see Table~\ref{t:obs}). Most of the epochs were obtained with an apodized Lyot coronagraph (\cite{Boccaletti2008}: the field mask in YJH has a radius of 92~mas, and  of 120~mas for the K-band coronagraph). Two of the observations (obtained in better observing conditions) were acquired without the coronagraph in order to study the very central region around the star. The use of the coronagraph allows a better contrast at separation larger than $\sim 0.1$~arcsec. For all data sets, the observations were acquired in pupil-stabilized mode. In addition to the science data, we acquired three kind of on-sky calibrations: (i) a flux calibration obtained by offsetting the star position by about 0.5 arcsec, that is, out of the coronagraphic mask (point spread function, PSF, calibration). (ii) An image acquired by imprinting a bidimensional sinusoidal pattern (waffle calibration) on the deformable mirror. The symmetric replicas of the stellar images obtained by this second calibration allow an accurate determination of the star centers even when the coronagraphic field mask is in place. These calibrations were obtained both before and after the science observation, and the  results were averaged. (iii) Finally, an empty field was observed at the end of the whole sequence to allow proper sky subtraction. This is relevant in particular for the K2 data sets.

Data were reduced to a 4D datacube (x, y, time, and $\lambda$) at the SPHERE Data Center in Grenoble (\cite{Delorme2017}) using the standard procedures in the SPHERE pipeline (DRH: \cite{Pavlov2008}) and special routines that recenter individual images using the satellite spot calibration, and correct for anamorphism, true north, and filter transmission. Faint structures can be detected in these images using differential imaging. Various differential imaging procedures were run on these data sets. We used here results obtained with a principal component analysis (PCA; see \cite{Soummer2012}) applied to the whole 4D datacubes, which combines both angular and spectral differential imaging in a single step (ASDI-PCA: see \cite{Mesa2015}). The PCA algorithm we used is the singular-value decomposition that generates the eigenvectors and eigenvalues that are used to reconstruct the original data. A principal components subset was used to generate an image with the quasi-static noise pattern that can then be subtracted from the original image. Clearly, the larger the number of principal components, the better the noise subtraction, but this also means that the signal from possible faint companion objects is cancelled out more strongly. Most of the results were obtained using 50 modes, but we also considered other numbers of modes (10, 25, 100, and 150 modes). To avoid spectrum distortion characteristics of the ASDI-PCA, photometry was obtained using a monochromatic PCA with only two modes for each spectral channel. Photometry was obtained with respect to the maximum of the PSF calibration corrected for the attenuation inherent to the PCA. The final step  of the procedure was to obtain signal-to-noise ratio (S/N) maps from the IFS images obtained by making a median over wavelengths. 

Finally, we also used the $Q_\Phi$\ image obtained by \cite{Pohl2017} for astrometry, reduced as described in that paper and in \cite{Ligi2017}. We note that this data set was obtained with the YJ field mask, whose radius is 72.5~mas.

\begin{table*}
\caption{Journal of observations}
\begin{centering}
\begin{tabular}{lccccccl}
\hline
\hline
JD       & Mode & nDIT$\times$DIT &   Angle & Seeing &  lim. cont  & coro & Ref\\
         &      &   (sec)          & (degree)  & (arcsec) &   (mag)       &      &\\
\hline
57145.   & Pol J & 3180 & Field & 0.90 & & YJ & \cite{Pohl2017} \\
57180.17 &  Y-J &   86$\times$64  &  ~45.82 &  1.57 &   13.13 & YJH & \cite{Ligi2017}\\        
57201.12 &  Y-H &   65$\times$64  &  ~36.42 &  1.00 &   13.52 & YJH & \cite{Ligi2017}\\          
57499.34 &  Y-J &   77$\times$64  &  144.62 &  1.88 &   13.06 & YJH & \cite{Ligi2017}\\          
57566.15 &  Y-H &  322$\times$2   &  147.33 &  0.67 &   13.64 &  no & \cite{Ligi2017}\\  
57873.30 &  Y-H &  192$\times$2   &  ~98.82 &  0.62 &   14.07 &  no & \cite{Ligi2017}\\
58288.19 &  Y-H &   48$\times$96  &  120.17 &  1.19 &   13.79 &  K  & This paper\\       
\hline
\end{tabular}
\end{centering}
\label{t:obs}
\end{table*}


\begin{figure*}
\centering
\includegraphics[width=18.8cm]{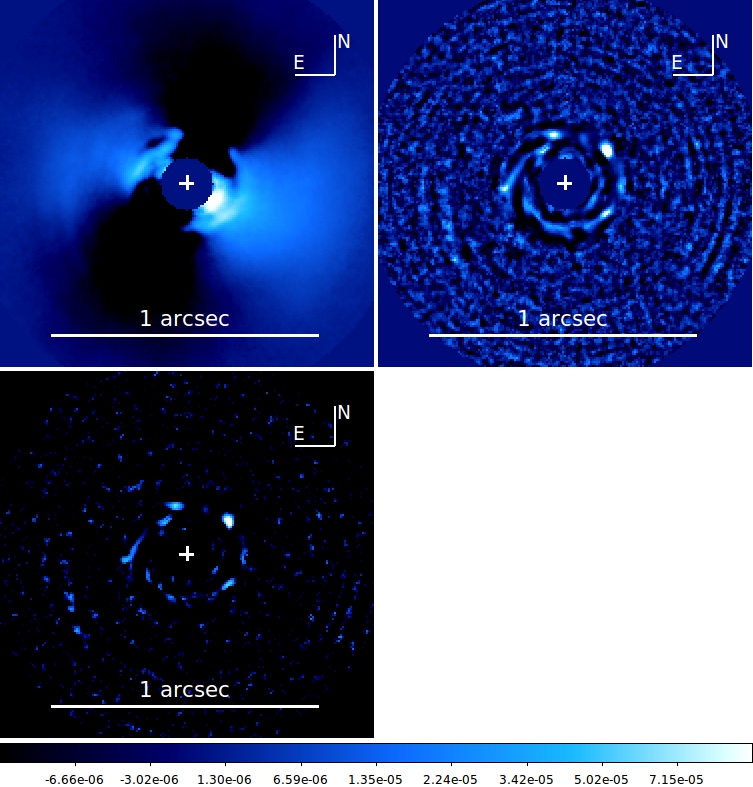}
\caption{Upper left panel: Median over time of the wavelength-collapsed images of HD169142 obtained with an ADI PCA algorithm, one mode per wavelength. Upper right panel: Image obtained by averaging the S/N maps for the individual epochs for IFS. The individual S/N maps are obtained from the ASDI PCA algorithm using 50 components and making a median of the wavelength (see \cite{Mesa2015}), and they were rotated for Keplerian motion to the last image before making the median. Lower panel: Coincidence image obtained from the same data set. In all panels, the solid line at the bottom of each panel represents 1 arcsec, and a white cross shows the position of the star. N is up and E to the left.}
\label{f:sum}
\end{figure*}

\section{Results}

The following discussion is based on the application of differential imaging algorithms that allow detecting faint structure that is not otherwise easily detectable in the images. The typical contrast of Ring 1 that we were able to measure using simple subtraction of a reference image is about $1.5\times 10^{-3}$. The structures we consider in this paper are more than an order of magnitude fainter. They represent small fluctuations of the signal that cannot be detected without differential imaging.

Figure~\ref{f:individual} shows the S/N maps obtained by applying the PCA ASDI algorithm to the IFS data for the individual epochs. The images have a linear scale from S/N=0 (dark) to S/N=5 (bright). These figures clearly show a similar pattern of bright spots, as well as a rotation of these spots with time. This suggests that a combination of the images that takes into account a Keplerian motion around the star should improve detection of the real pattern present in the data. The full solution is quite complex, leaving many free parameters, and may be attempted using an approach such as that considered by K-stacker (\cite{Nowak2018}). However, a simplified approach that greatly reduces the number of free parameters is to assume that the system is seen face-on and that the orbits are circular: if the distance is known, the only free parameter is the stellar mass. This appears to be a reasonable approximation for disk-related features around HD~169142 because in this case, we only consider a fraction of the orbit. On the other hand, observations spread over a few years enable separating static features that are due to radiative transfer effects from scattered-light fluctuations that are due to moving clouds or sub-stellar objects.

The upper panels of Figure~\ref{f:sum} show images of HD169142 obtained by combining the six individual images, assuming the distance given by GAIA DR2, a mass of 1.7~M$_\odot$\ (see below), and circular orbits.

\subsection{Coincidence images}

To improve our ability of discerning faint signals, we combined data from different epochs using a coincidence map (see the lower panel of Figure~\ref{f:sum}). The principle of this coincidence map is to start with S/N maps for individual epochs. We used S/N maps after correcting for the small-number-statistics effect using the formula by \cite{Mawet2014}. The maps were then multiplied by each other. The S/N maps average to zero, with both positive and negative values for individual pixels. Of course, this may result in a false-positive signal for an even number of negative signals in the individual S/N maps. To avoid this problem, we arbitrarily set to negative the result for a given pixel when the signal for that pixel was negative for at least one epoch. Of course, this is not a realistic flux map: the aim is merely to identify consistent signals throughout all individual images.

To consider the orbital motion around the star over the three years covered by our observations, we divided the field into 65 rings, each one 2 IFS pixels wide (15 mas, i.e., about 1.8 au at the distance of HD~169142). For each ring, we rotated the S/N maps obtained at different epochs with respect to the first reference image according to Kepler's third law. When the distance to the star was fixed, the only free parameter remaining in this model is the (dynamical) stellar mass. If there is a companion orbiting the star, the signal is maximized for a value of the mass that, if the assumptions made (circular motion seen face on) are correct, is the dynamical mass of the star. If these hypotheses are not correct, the estimate of the mass is incorrect, by a value that depends on the real orbital parameters. We adopted a mass of 1.7~$M_\odot$, the GAIA parallax, and clockwise rotation, as indicated by the analysis of motion of disk features in \cite{Ligi2017}; see also \cite{Macias2017}. Figure~\ref{f:sum} shows the coincidence map and a mean of the S/N maps for the six epochs for this value of the mass.

\begin{figure}
\centering
\includegraphics[width=\columnwidth]{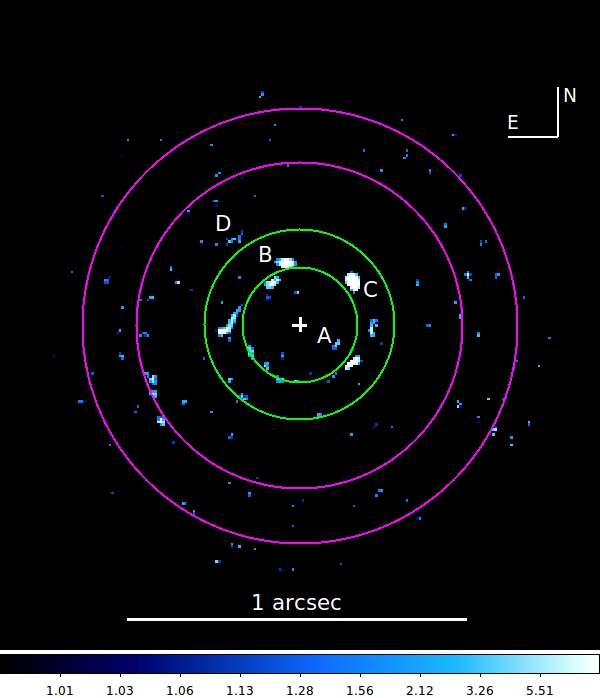}
\caption{Same as the lower panel of Fig~\ref{f:sum}, but showing the edges of the two disk rings (Ring 1 in green, Ring 2 in magenta). The ring edges are drawn according to \cite{Fedele2017}. The blobs are labeled. The white solid line represents 1 arcsec, and the white cross shows the position of the star. N is up and E to the left.}
\label{f:rings}
\end{figure}

\subsection{Blob detection}

A quite large number of blobs can be found around HD~169142. Several of them are found consistently in all individual images and are also visible in the J-band $Q_\Phi$ image seen in Figure~\ref{f:pdi}; some of them have been identified and discussed by \cite{Ligi2017}.  We fixed our attention on four of them (see Figure~\ref{f:pdi} and Figure~\ref{f:rings} for their definition). The two brightest blobs (B and C) are within Ring 1 and have been identified by \cite{Ligi2017}; they called them blobs A and B, respectively. Our blob A is closer to the star than Ring 1. Blob D is between Rings 1 and 2. All of these blobs appear to be slightly extended. We verified in the individual images that this is not an artifact caused by combining individual images. For instance, when we consider the best set of data (the last set from June 2018), the FWHM of blobs B and C can be measured with reasonable accuracy at about 40 mas, which is significantly larger than expected for a point source at this separation (about 26 mas, after applying differential imaging). To better estimate the physical size of the blobs, we compared the FWHM measured in our differential images with the FWHMs obtained for fake blobs that are the result of convolving Gaussian profiles with the observed PSF inserted into the images at the same separation but at a different position angle, and processed through the same differential imaging procedure. We repeated this procedure for the images obtained considering 50 modes (best image for detection) and with a less aggressive image where only 25 modes were considered,  which better conserves the original shape of the blobs. In this way, we found that the FWHM of blobs B and C is the same as that of fake Gaussian blobs with an intrinsic FWHM of 42 and 30 mas for blobs B and C, respectively. However, these are average values for tangential and radial profiles (with respect to the star): both blobs appear elongated in the tangential (rotation) direction with axis ratios of 1.4 (blob B) and 1.9 (blob C)\footnote{This is not as obvious from a simple visual inspection of the images because the ADI processing that is implicit in the PCA-ASDI procedure we used deforms the images.}. The uncertainty on this size estimate is of about 7 mas, as obtained by comparing the results obtained in individual images. This size corresponds to $\sim 4-5$~au, with an uncertainty of about 1~au. This result should be considered with some caution because the light distribution of the blobs might be not well reproduced by Gaussians.

\begin{figure}
\centering
\includegraphics[width=\columnwidth]{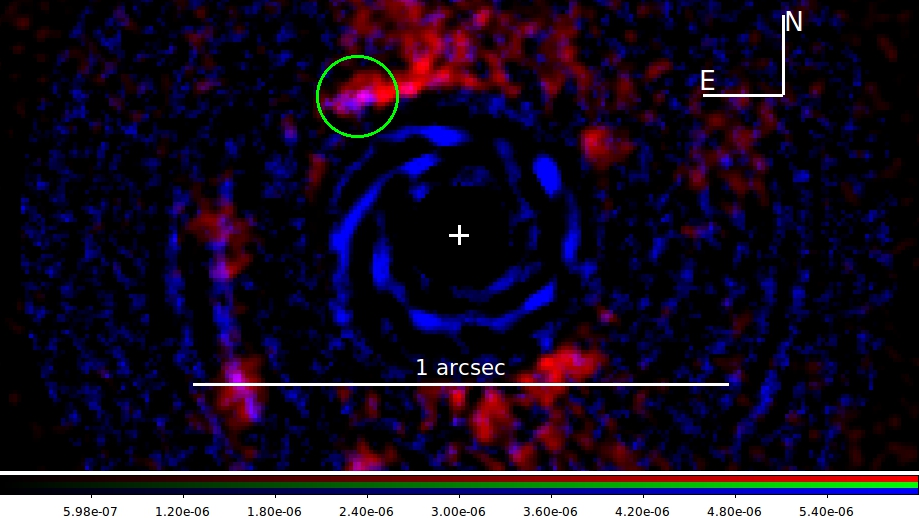}
\caption{Zoom of a two-color image of the region around blob D. This image was constructed using the K2 observation of JD=2458288.19 (red) and the weighted sum of all the IFS images (collapsed against wavelength) and rotated for a Keplerian motion assuming that the star has a mass of 1.7~M$_\odot$ (blue). This last image is for the same epoch as the K2 observation. For clarity, the region within 0.28 arcsec from the star (i.e., within the outer edge of Ring 1) was masked in the K2-band image. The green circle is centered on the position of the blob measured in the K2 image. We note the different aspect and small offset between the position of the blob in the K2 image with respect to that at shorter wavelength. The white solid line represents 1 arcsec. N is up and E to the left. The white cross marks the position of the star.}
\label{f:blobd}
\end{figure}

In Figure~\ref{f:blobd} we show a zoom of the region around blob D in a two-color image. The blue structures visible in the image are obtained through IFS in the Y, J, and H bands. They could be interpreted as stellar light scattered by a (dusty) spiral structure around a protoplanet that is accreting material funnelled through the spiral arm from the disk. This interpretation agrees with the detection in the $Q_\Phi$\ image (see Figure~\ref{f:pdi}). In the same image, the red structures are obtained through IRDIS observing in the K2 band. In particular, the structure in the green circle that appears to be much more similar to a point source might indicate a planetary photosphere. The position of the blob in the K2 image is Sep=332 mas, PA=34.9 degree, which is not the photocenter at shorter wavelengths. Even if this interpretation is speculative (there are other structures in this image that we consider as noise), various circumstantial arguments discussed below possibly support it. We return to this point in the next section.

There is of course some probability that these detections are spurious. In order to estimate the false-alarm probability (FAP), we proceeded as follows. First, we fixed the stellar mass at the value given by fitting isochrones (1.7~$M_\odot$). With this assumption, the prediction for the orbital motion is fully independent of the SHINE data set. We derotated the individual images to the same epoch using the same approach as described above (ring by ring). We searched for signals in the final coincidence data set using the FIND procedure in IDL. We recovered the detection of the candidate. We ordered the different epochs according to the value of the S/N at the candidate positions (separately for each candidate). We then used binomial statistics on the remaining epochs (i.e., excluding the reference with the highest S/N), considering as number of trials the number of pixels with an S/N higher than the S/N measured in the candidate position in the image giving the highest S/N value at this position. To estimate the probability in the binomial statistics, we considered the product $\prod_c$\ of the S/N rankings in the pixel corresponding to the candidate position in the remaining images, and compared this product to a similar product $\prod_r$ obtained from random extractions. We repeated the random extraction $10^7$ times, and assumed that the probability of success is given by the fraction of cases where $\prod_r < \prod_c$.

With this approach, we obtained the FAP values listed in the second column of Table~\ref{t:blobrot}; values for blobs B, C, and D are highly significant.


\subsection{Blob astrometry}

\begin{figure*}
\begin{tabular}{cc}
\centering
\includegraphics[width=8truecm]{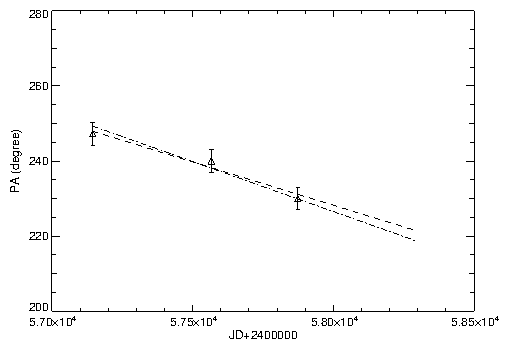}&
\includegraphics[width=8truecm]{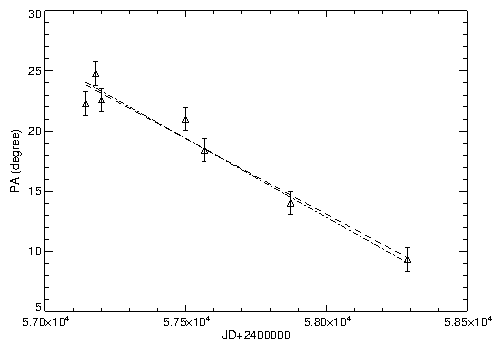}\\
\includegraphics[width=8truecm]{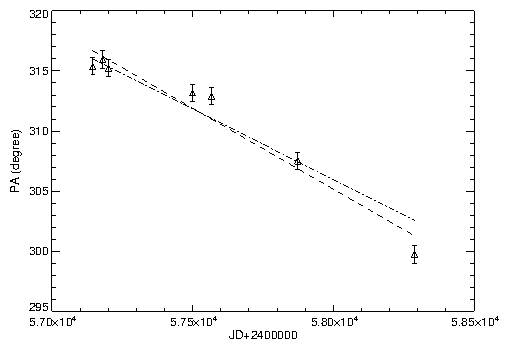}&
\includegraphics[width=8truecm]{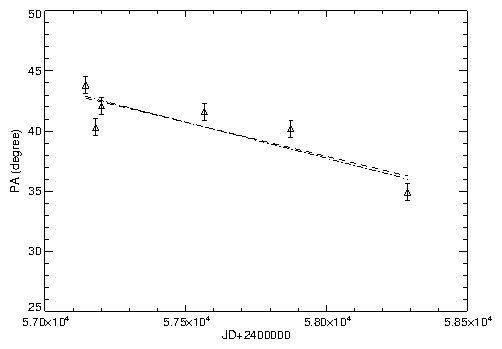}\\
\end{tabular}
\caption{Variation in PA of the blobs with time. Upper left panel: blob A. Upper right panel: blob B. Lower left panel: blob C. Lower right panel: blob D. The dashed lines are best-fit lines through the points. Dash-dotted lines are predictions for circular orbits assuming a mass of 1.85~M$_\odot$\ for the star.}
\label{f:astro}
\end{figure*}

\begin{figure}
    \centering
    \includegraphics[width=\columnwidth]{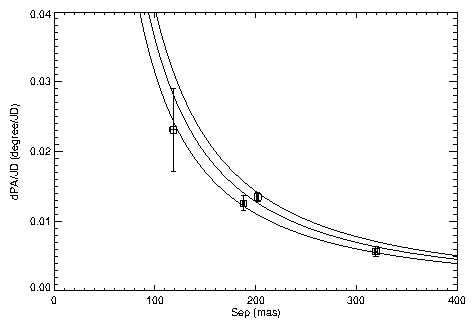}
    \caption{Run of the angular speed as a function of separation for the four blobs around HD169142; these values are on the disk plane. Overimposed we show predictions for circular Keplerian motion for three different values for the stellar mass (1.5, 2.0, and 2.5 M$_\odot$).}
    \label{f:kepler}
\end{figure}

All these blobs rotate around the star. This can be shown by measuring their position in the individual images (see Table~\ref{t:blobastro}).  We used the IDL FIND algorithm that uses marginal Gaussian distributions to measure the position of the spot centers in the ASDI 50 components images obtained at the various epochs. In addition, we also measured the blob positions in the polarimetric image. We found that the rotational speed decreases with separation (see Table~\ref{t:blobrot} and Figure~\ref{f:astro}), as expected for Keplerian motions. Figure~\ref{f:kepler} shows the run of the angular speed as a function of separation for the four blobs around HD169142; these values are on the disk plane. Overimposed are predictions for circular Keplerian motion for three different values for the mass of the star (1.5, 2.0, and 2.5 M$_\odot$). 

When we interpret the observed angular motion as Keplerian circular orbits in the disk plane, we can determine the mass of HD~169142. When we use the three blobs B, C, and D, the mass of the star is $1.85\pm 0.25$~M$_\odot$ (we did not use blob A here because it has too few astrometric points). When we add the uncertainties that are due to parallax  (0.05 M$_\odot$) and disk inclination (0.09 M$_\odot$), the result is $1.85\pm 0.27$~M$_\odot$. 

The mass estimated by this procedure might be underestimated because the photocenter of the blobs might be closer to the star than their center of mass and the mass determination depends on the cube of the separation. In particular, as noted above, we may interpret blob D as the accretion flows on a planet along a spiral arm (see the next section); in this case, the photocenter is dominated by the leading arm, which is at about 313 mas from the star, while the trailing arm is at 347 mas when it is deprojected on the disk plane. The putative planet would be in the middle of the two arms, that is, at 330 mas from the star, yielding a mass estimate that is $\sim 10$\% higher than listed in Table~\ref{t:blobrot}. We note that the separation measured in the K2 band agrees very well with this interpretation.

The mass determined from blob motion is slightly higher than but in agreement within the error bars with the mass that fits photometry. To show this, we determined the stellar mass by minimizing the $\chi^2$ with respect to the main-sequence values considered by \cite{Pecaut2013}. We considered the GAIA DR2 parallax and included an absorption term $A_V$ multiplied for the reddening relation by \cite{Cardelli1989}. We also left free the ratio between the stellar and the main-sequence radius. The best match is with an F0V star ($T_{\rm eff}=7220$~K), with $A_V=0.25$~mag and a radius that is 0.97 times the radius of the main-sequence star. According to \cite{Pecaut2013}, the mass of an F0V star is 1.59~M$_\odot$. This spectral type compares quite well with the most recent determinations (A7V: \cite{Dent2013}; A9V: \cite{Vieira2003}; F0V: \cite{Paunzen2001}; F1V: \cite{Murphy2015}) and with the temperature determined by GAIA ($T_{\rm eff}=7320\pm 150$~K), but it is much later than the B9V spectra type proposed by \cite{Wright2003}. 

For comparison, other determinations of the mass of HD~169142 are 2.0~M$_\odot$\ (\cite{Manoj2005}), 2.28~M$_\odot$\ (\cite{Maaskant2013}), 1.8~M$_\odot$\ (\cite{Salyk2013}), and 2.0~M$_\odot$\ (\cite{Vioque2018}). The mass adopted by \cite{Ligi2017} is 1.7~M$_\odot$. We note that these values were obtained assuming distances different from the distance given by GAIA DR2: for instance, \cite{Maaskant2013} adopted a distance of 145 pc, which is 27\% larger than the GAIA DR2 value considered here. On the other hand, the value used by \cite{Ligi2017} was taken from GAIA DR1 and it is only 3\% longer than that from GAIA DR2.
Hereafter, we adopt a mass  of 1.7~M$_\odot$ for HD~169142, which is the same value as was considered by \cite{Ligi2017}.

We also note that the projected rotational velocity of the star $V~\sin{i}=50.3\pm 0.8$~km/s determined from the HARPS spectra (see Appendix) is high when we consider that the star is likely seen close to the pole. This value agrees quite well with literature values ($V~\sin{i}=55\pm 2$~km/s: \cite{Dunkin1997a, Dunkin1997b}). When we assume that the stellar rotation is aligned with the disk, the equatorial rotational velocity is 224~km/s, which is at the upper edge of the distribution for F0 stars. For a discussion, see \cite{Grady2007}.


\begin{table*}
\caption{Blob astrometry}
\begin{centering}
\begin{tabular}{lcccccccc}
\hline
\hline
JD       &\multicolumn{2}{c}{Blob A}&\multicolumn{2}{c}{Blob B}&\multicolumn{2}{c}{Blob C}&\multicolumn{2}{c}{Blob D}\\
+2400000 & Sep & PA & Sep & PA & Sep & PA & Sep & PA \\
         & (mas) & (degree) & (mas) & (degree) & (mas) & (degree) & (mas) & (degree) \\
\hline
57145.   & $106\pm 6$ & $247\pm 3$ & $185.4\pm 4.0$ & $22.3\pm 1.0$ & $192.7\pm 4.0$ & $315.8\pm 2.0$ & $315.8\pm 4.0$ & $43.8\pm 0.7$ \\ 
57180.17 &            &            & $194.0\pm 3.2$ & $24.8\pm 1.0$ & $197.9\pm 2.5$ & $316.0\pm 0.7$ & $313.9\pm 4.0$ & $40.3\pm 0.7$ \\
57201.12 &            &            & $188.3\pm 3.2$ & $22.6\pm 1.0$ & $202.8\pm 2.5$ & $315.2\pm 0.7$ & $314.8\pm 4.0$ & $42.1\pm 0.7$ \\
57499.34 &            &            & $187.7\pm 3.2$ & $21.0\pm 1.0$ & $197.6\pm 2.5$ & $313.2\pm 0.7$ &                &               \\
57566.15 & $125\pm 6$ & $240\pm 3$ & $188.7\pm 3.2$ & $18.4\pm 1.0$ & $203.7\pm 2.5$ & $312.9\pm 0.7$ & $315.5\pm 4.0$ & $41.6\pm 0.7$ \\
57873.30 & $117\pm 6$ & $230\pm 3$ & $184.6\pm 3.2$ & $14.0\pm 1.0$ & $200.1\pm 2.5$ & $307.5\pm 0.7$ & $319.2\pm 4.0$ & $40.2\pm 0.7$ \\
58288.19 &            &            & $189.7\pm 3.2$ & $~9.8\pm 1.0$ & $200.1\pm 2.5$ & $299.7\pm 0.7$ & $315.6\pm 4.0$ & $34.9\pm 0.7$ \\
\hline
\end{tabular}
\end{centering}
\label{t:blobastro}
\end{table*}

\begin{table*}
\caption{Blob rotation}
\begin{centering}
\begin{tabular}{lcccccccc}
\hline
\hline
Blob     &  FAP & a  &  a   & Period   & Period       & Rot. speed &  Mass      & Remark \\      
         &      &    &      & Computed & Observed     &            &            &        \\
         &      & (mas) &  (au)  &  (yr)      & (yr)           & (deg/yr)     & (M$_\odot$)  &        \\
\hline

A   &   0.02  &  118 & 13.5 &   36.2   &  42.7$\pm$5.6  & -11.9$\pm$2.2 
& 1.60$\pm$0.98     & pol and nocoro images \\
B   & $<1E-7$ &  188 & 21.4 &   72.5   &  78.3$\pm$5.3  & -5.04$\pm$0.38 & 2.30$\pm$0.29 & \\
C   & $<1E-7$ &  202 & 23.1 &   80.9   &  73.0$\pm$4.3  & -4.48$\pm$0.25 & 1.60$\pm$0.23 & \\
D   &  4E-6   &  319 & 36.4 &  160.5   & 173.8$\pm$20.1 & -2.08$\pm$0.25 & 1.34$\pm$0.40 & \\
\hline
\end{tabular}
\\
Note: Semi-major axis a is obtained assuming circular orbits on the disk plane; the computed period is for a mass of 1.87~$M\odot$;  the observed period is estimated from the angular speed on the disk plane; the mass is determined using Kepler's third law; the uncertainty here is due to the errors in the angular speed.
\end{centering}
\label{t:blobrot}
\end{table*}

\subsection{Blob photometry}

\begin{figure}
\centering
\includegraphics[width=\columnwidth]{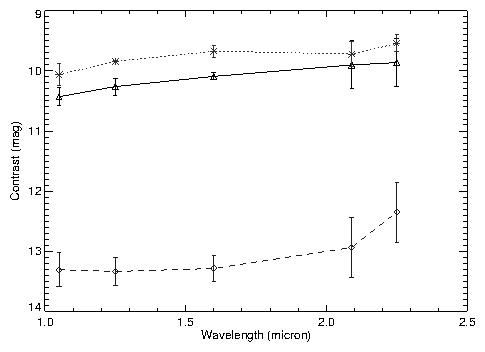}
\caption{Contrast of blobs as a function of wavelength. Blob B: Asterisks and dotted line. Blob C: Triangles and solid line. Blob D: Diamonds and dashed line.}
\label{f:contrast}
\end{figure}

We measured the magnitude of the sources in various bands by weighting the results obtained from the different epochs according to the quality of the images. The magnitudes refer to a $3\times 3$ pixel area centered on each object and are obtained by comparison with those of simulated planets inserted into the image at 0.2 and 0.3 arcsec from the star and run through the same differential imaging algorithm. The underlying assumption is that the blobs are point sources, while they are likely slightly extended. These results should then be taken with caution. Using the fake blob procedure described in Section 3.2, we estimated that the brightness is underestimated by about a factor of $\sim 2.8$~because of this effect for blobs B and C, that is, these blobs are likely $\sim 1.1$ magnitude brighter than estimated when we assume that they are point sources. The effect is likely slightly smaller for blob D because it is farther away from the star. We summarize the results in Table~\ref{t:blobphot}; error bars are the standard deviation of the mean of the results obtained at different epochs. All the blobs have a rather flat, only slightly reddish contrast with respect to the star (see Figure~\ref{f:contrast}). Results are consistent with stellar light scattered by grains with a size on the order of a micron or smaller if stellar light is extinguished between the star and blobs or the blobs and us. Under the hypothesis (not demonstrated) that they are optically thick, the albedo required to reproduce observations of blobs A, B, and C is about 0.1.

\begin{figure}
\centering
\includegraphics[width=\columnwidth]{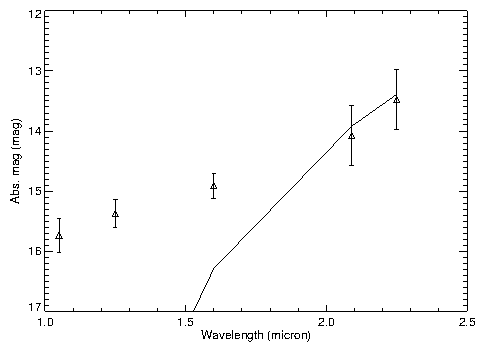}
\caption{Absolute magnitude  of blob D in various bands (diamonds). The solid line is the prediction for a 3~M$_J$, 5~Myr old planet using dusty isochrones by \cite{Allard2001}}
\label{f:photo_b}
\end{figure}

Blob D is about two magnitudes fainter than expected from this consideration, suggesting that either it receives less light from the star (e.g., because of absorption by Ring 1) or it is not optically thick. For this blob, we obtained contrasts of $13.31\pm 0.28$~mag in Y, $13.34\pm 0.23$~mag in J band, $13.29\pm 0.21$~mag in H band, $12.94\pm 0.5$~mag in  K1 band, and $12.35\pm 0.5$~mag in K2 band (the last two values being obtained from the IRDIS data set). As expected, the object is beyond the 5$\sigma$\ contrast limit in each individual image. However, we expect a detection with an S/N in the range from 2.3 to 3.7 in the individual images, and at an S/N$\sim 6$\ in the combination of the images. It is then not surprising that we detected it only by combining them. While error bars are quite large, the absolute K1 and K2 magnitude of $14.07\pm 0.50$~mag and $13.48\pm 0.50$~mag corresponds to a $\sim 3$~M$_J$ object using dusty isochrones (\cite{Allard2001}) with an age of 5 Myr, which is at the lower edge of the age range according to \cite{Blondel2006} and \cite{Manoj2007}. This model has an effective temperature of about 1260 K (see Figure~\ref{f:photo_b}). Of course, the mass estimated from photometry depends on the model, the age used to derive it and the possible extinction, and it assumes that the object is in hydrostatic equilibrium, which may be incorrect for a very young planet. This result is then highly uncertain.

\begin{table*}
\caption{Blob photometry. These values are obtained assuming that the blobs are point sources; they may be as much as 1.1 mag brighter if their extension is taken into account}
\begin{centering}
\begin{tabular}{lccccc}
\hline
\hline
Blob & \multicolumn{5}{c}{Contrast (in magnitudes) } \\
     &       Y      &     J       &      H      &     K1      &      K2      \\
\hline
A    &      9.05    &    8.73     &    9.02\\
B    & 10.06$\pm$0.19 & 9.84$\pm$0.01 & 9.67$\pm$0.10 & 9.72$\pm$0.21 & 9.54$\pm$0.14  \\
C    & 10.43$\pm$0.15 &10.26$\pm$0.14 &10.09$\pm$0.06 & 9.90$\pm$0.40 & 9.86$\pm$0.40  \\
D    & 13.31$\pm$0.28 &13.34$\pm$0.23 &13.29$\pm$0.21 &12.94$\pm$0.50 &12.35$\pm$0.50  \\
\hline
\end{tabular}
\end{centering}
\label{t:blobphot}
\end{table*}

\subsection{Comparison with previous detection claims}

We note that none of these blobs coincides with either the sub-stellar companions proposed by \cite{Biller2014} and \cite{Reggiani2014}, nor with the structure observed by \cite{Osorio2014}. More in detail, after taking into account their motion (see Section 3.3), the expected position angles for blobs A, B, and C at the observation epochs of Biller et al. and Reggiani et al. (both acquired at an epoch about 2013.5), are  276, 34, and 324 degree, respectively (blob D is much farther away from the star). For comparison, the object of Biller et al. is at PA=$0\pm 14$~degree (separation of $110\pm 30$\  mas) and the object of  Reggiani et al. is at PA=$7.4\pm 11.3$ degree (separation of $156\pm 32$~mas). In addition, the objects proposed by Biller et al. and Reggiani et al., with a contrast of $\Delta L\sim 6.5$, are brighter than our blobs B and C, even after the finite-size correction is taken into account, see Section 3.4. However, the object proposed by Reggiani et al. might be the combination of blobs B and C, within the errors of their astrometry; the combination of their luminosity is also not that far from the value of Reggiani et al. We note that the resolution of their observation is lower than ours because they observed at much longer wavelength, and their object appears elongated (in the E-W direction, i.e., the direction expected at the epoch of their observation) in their published image beyond the diffraction limit.

On the other hand, the inner and brighter object detected by Biller et al. is too close to the star to coincide with any of the objects we observed, while a fainter object they found might be blob B, as discussed by \cite{Ligi2017}. However, when we examine the image published by Biller et al, it seems that the two brightest sources have a relative separation and orientation that coincides with those of blobs B and C. In this case, the fainter object should be blob B (as discussed in \cite{Ligi2017}) and the brighter object might coincide with our blob C. Of course, this would require that the stellar position in their images does not correspond with the position assumed in their paper.

\section{Spiral arms}

\begin{figure}
\centering
\includegraphics[width=8.8truecm]{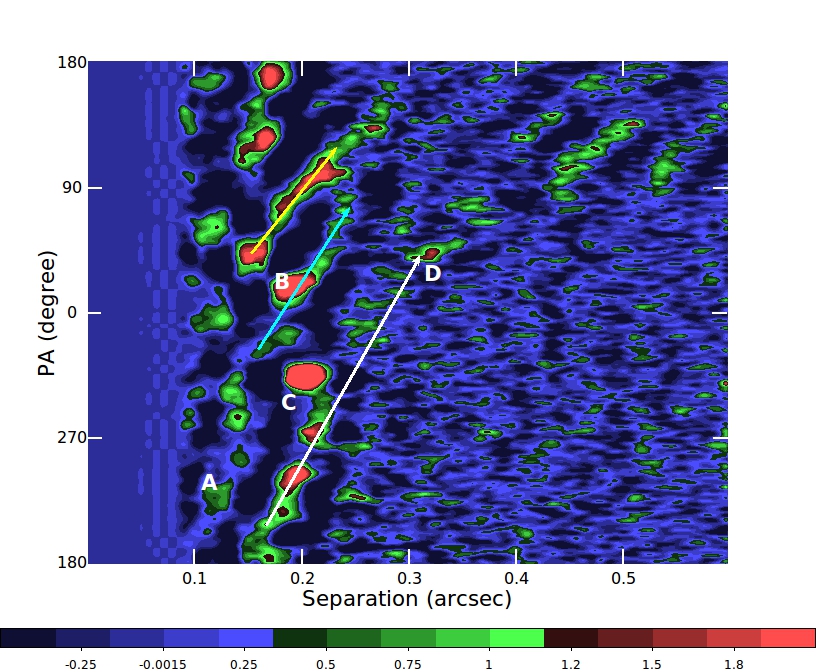}
\caption{Median over time of the individual S/N maps in polar coordinates. Each image has been rotated to the last image for the rotation angle of blob D before the median was made. Arrows mark the location of the primary (white), secondary (cyan), and tertiary arms (yellow). The location of the blobs is marked.}
\label{f:polar}
\end{figure}

\begin{figure}
\centering
\includegraphics[width=\columnwidth]{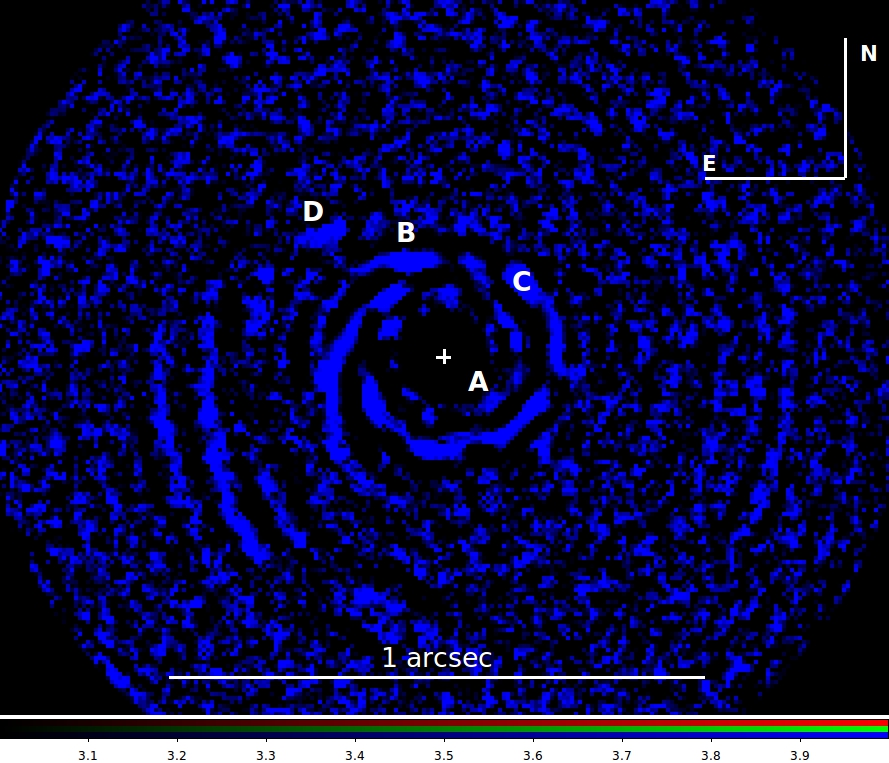}
\includegraphics[width=\columnwidth]{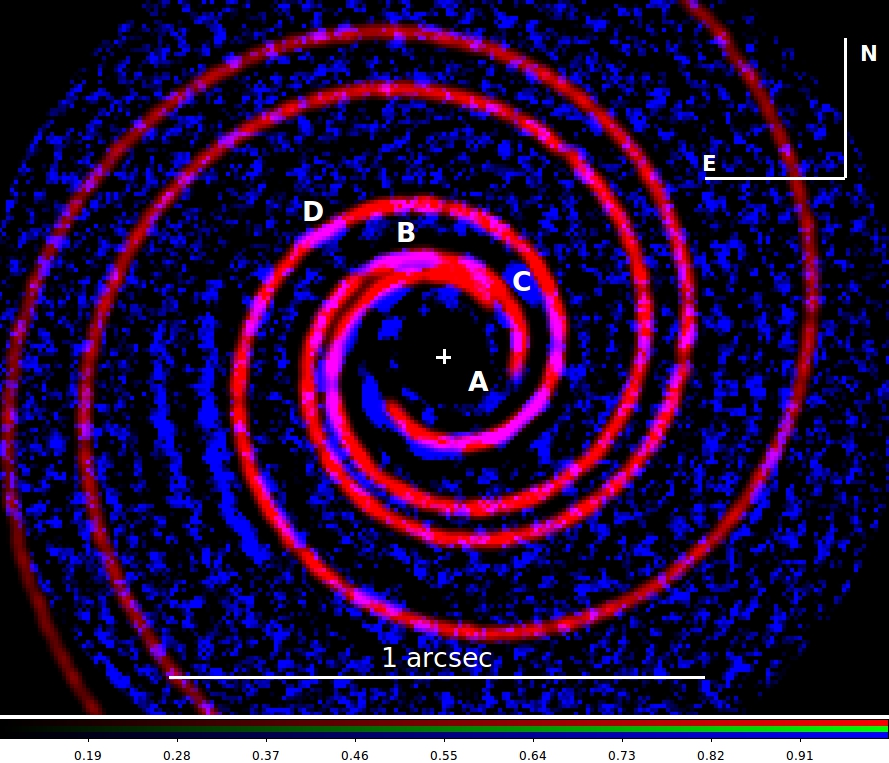}
\caption{Upper panel: Median over time of the individual S/N maps; each image has been rotated to the last image for the rotation angle of blob D before the median was made. Lower panel: Same as the upper panel, with a three-arm spiral design overplotted. The putative planet is at the position of blob D. In both panels, N is up and E to the left, and a white cross marks the stellar position.}
\label{f:spirals}
\end{figure}

Most bright features in the coincidence images (including blobs B and D) can be reproduced by a three-arm spiral design (see below), while blob C differs slightly. A similar three-arm structure is predicted by models for not very massive companions (\cite{Fung2015}) and it has been observed around other stars (see, e.g., the case of MWC758 recently published by \cite{Reggiani2018}). While it is not at all obvious that spiral arms indicate a planet (see, e.g., \cite{Dong2018}), we may interpret it as due to a planet in the location of blob D. The structure of this blob appears to resemble the structure expected for an accreting object with a leading and a trailing arm. If this hypothesis were correct, the radial separation between leading (sep=310 mas) and trailing (sep=343 mas) arms should be about twice the Hill radius (see \cite{Machida2010}); the Hill radius would then be $16.5\pm 5$~mas and the planet mass in the range 0.25-1.6~M$_J$, which is lower than the mass estimated with DUSTY isochrones. Because the Hill radius is not accurately estimated and the dependence of the mass on the Hill radius is strong, the error on the planetary mass is quite large. The photosphere of such an object would be too faint for detection in YJH, while it might be compatible with detection in K1 and K2 bands.

\subsection{Separation of spiral arms}

We may also estimate the mass of the object exciting the spiral design observed in Ring 1 by different criteria, using the calibration by \cite{Fung2015} (their eq. (9)). After transforming into a polar coordinates system (see Figure~\ref{f:polar}), we could identify the three spiral arms, which we may call primary, secondary, and tertiary, following the approach of Fung \& Dong. The view in Cartesian coordinates is given in Figure~\ref{f:spirals}. Position angle and separation of the arms in some reference positions are given in Table~\ref{t:spiral}.

\begin{table}
\caption{Spiral position.}
\begin{tabular}{lccc}
\hline
\hline
      &    PA   &     PA     &     PA    \\
Sep   & Primary &  Secondary &  Tertiary \\
(mas) & (degree)  & (degree)   &   (degree)  \\
\hline
157   &  196.7  &   321.0    &    38.9   \\
172   &  203.0  &   335.9    &    50.9   \\
194   &  244.2  &    16.0    &    79.0   \\
209   &  268.3  &    28.0    &    91.6   \\
\hline
pitch &   15.3  &    16.3    &    20.8   \\
\hline
\end{tabular}
\label{t:spiral}
\end{table}

These arms may be density waves excited by a a planet at the location of blob D, which is indeed along the primary arm of the spiral design: the predicted PA at the separation of blob D, 334 mas, is $36\pm 6$~degree, in very good agreement with the observed value of $\sim 35$~degree (as measured in the K2 band).

The phase difference between the primary and secondary arm ($127.2\pm 3.1$~degree) can be used to estimate the mass of the planet exciting the spiral design, using the calibration by Fung \& Dong. We obtain a mass ratio of $q=0.0030\pm 0.0004$, which translates into a mass of $M_p=5.1\pm 1.1$~M$_J$, adopting the stellar mass derived above. The phase difference between secondary and tertiary arms ($69.9\pm 3.9$~degree) agrees with the expectations by Fung \& Dong given the pitch angle and the expected ratio for resonances 1:2 and 1:3.

\subsection{Pitch angle}

The pitch angle is the angle between a spiral arm and the tangent to a circle at the same distance from the star. \cite{Zhu2015} showed that the pitch angle can be used to estimate the mass of the planet exciting the spiral design. We measured the pitch angle at a separation of 183 mas to be $17.5\pm 1.7$~degree. This separation is about $r/r_p=0.55$. This value agrees with the results they obtained from their simulations for a mass ratio of $q=0.006$, supporting the mass determination obtained from the separation of primary and secondary spiral arm; moreover, the larger pitch for the tertiary arm agrees with expectations from models.

\subsection{Disk gap}

Using the relation by \cite{Kanagawa2016}, we expect that there is a planet at $\sim 0.36$~arcsec from the star with a mass ratio with respect to the star of $q=2.1\times 10^{-3}\,(W/R_p)^2\,(h_p/0.05~R_p)^{1.5}\,(\alpha/10^{-3})^{0.5}$, where $h_p/R_p$ is the disk thickness and $\alpha$\ is the disk viscosity. For $R_p=0.36$~arcsec, $W=0.2$~arcsec, $h_p/R_p=0.05$, and $\alpha=1E-3$, a value of $q=0.00044$\ is obtained, which means a planet of 0.75~M$_J$. 

\cite{Dong2017} considered the case of HD~169142 and concluded for a value of $q^2/\alpha=1.1E-4$\ for $R_p=0.37$~arcsec, $W=0.17$\ arcsec, and $h_p/R_p$=0.079. For $\alpha=1E-3$, their formula implies $q=0.00033,$\  which suggests a 0.56~M$_J$\ planet. We note that the formula by Dong \& Fung produces planets that are smaller by a factor of 2.6 with respect to that by Kanagawa et al.; however, the value they suggest for $h_p/R_p$\ is higher than considered above. The value considered by Dong \& Fung is similar to the value obtained by \cite{Fedele2017} by modeling the ALMA observations ($h_p/R_p$=0.07).

There are considerable uncertainties in these formulas that are due to the exact values to be adopted for $R_p$, $W$, $h_p/R_p$, $\alpha$, and the difference of a factor of 2.5 in the constant factor. While a mass around 1~M$_J$\ seems favored, we cannot exclude values different by as  much as an order of magnitude. We conclude that a planet with about one Jupiter mass likely causes the gap seen in HD169142, but its mass is not yet well defined from the gap alone.

\begin{table}
\caption{Putative planet mass (sep=335 mas, PA=35 degree at JD=58288.19)}
\begin{tabular}{lcc}
\hline
\hline
Method          &   M$_J$   & Remark \\
\hline
Photometry      &    3      & Age dependent \\
Hill radius     &  0.25-1.6 & \\
Spiral arm separation  &  4.0-6.2  & \\
Pitch angle     &    6      & \\
Disk gap        &  0.06-6   & \\
\hline
\end{tabular}
\label{t:mass}
\end{table}

\subsection{Summary of mass determination}

A summary of the mass determinations is given in Table~\ref{t:mass}. All these estimates are quite uncertain. The higher values are given by the spiral arm parameters. If we make an harmonic mean of the various estimates, we would conclude for a planet with a mass of $2.2_{-0.9}^{+1.4}$~M$_J$. This mass seems lower than what we can detect with our SPHERE images (about 3~M$_J$\ from photometry), but is within the error bar. This value is also within the range 1-10~M$_J$\ suggested by \cite{Fedele2017} to justify the dust cavity observed with ALMA between rings 1 and 2, and it is on the same order as the missing mass in the disk within the gap, as given by their disk model (4.3~M$_J$).
For comparison, we note that if we were to try to interpret the spiral arms of MWC~758 (\cite{Reggiani2018}) using the same approach, we would conclude for a more massive faint companion because in that case the separation between the primary and secondary arm is much closer to 180 degree.

\section{Conclusion}

We performed an analysis of faint structures around HD~169142 that are persistent among several data sets obtained with SPHERE and analyzed them using differential image techniques. We found a number of blobs that rotate around the star as well as spiral arms. These structures represent small fluctuations of the overall disk structure around this star. The blobs are found to consistently rotate around the star with Keplerian circular motion.

Although we cannot exclude other hypotheses, blob D might correspond to a low-mass ($\sim 1-4$~M$_J$, best guess of 2.2~M$_J$), 5 Myr old, and still-accreting planet at about 335 mas (38 au) from the star, causing the gap between Rings 1 and 2 and exciting the spiral arm design observed within Ring 1. The separation between the outer edge of Ring 1 and blob D is 55 mas, which is about twice the proposed value for the Hill radius of the planet. There is a clear excess of flux at short wavelengths with respect to the flux expected for a planetary photosphere (see Figure~\ref{f:photo_b}). In our proposed scenario, the planetary photosphere is not detected in YJH band, where we only see the accreting material fueling through the spiral arm and reflecting star light (consistent with its detection in the $Q_\Phi$\ image), while it might have been detected in the K1 and K2 bands. A planet of 2.2~$M_J$\ at 335~mas (38~au) from HD~169142 would have a Hill radius of about 25 mas (3.2 au). A disk around such an object would have an FWHM slightly larger than the resolution limit of SPHERE and may well reflect some $10^{-5}$ of the stellar light, which is required to justify the flux observed in the YJH bands. On the other hand, it is also possible that no other planet exists, and we merely observe a dust cloud. Detection of a planet could be confirmed by observations in the L' band. According to the AMES-dusty isochrones (\cite{Allard2001}), a 5 Myr old planet of 2.2~$M_J$\ should have an absolute L' magnitude of $\sim 11$~mag. The contrast in the L' band should then be of 10.1 mag, which is 3.7 mag fainter than the objects proposed by \cite{Biller2014} and \cite{Reggiani2014} and likely too faint for a detection in their data set. However, a future deeper data set can solve this issue.

The location of blob B (and C to a lesser extent) suggests at first sight that the blobs might be related to the secondary and tertiary spiral arms (see, e.g., \cite{Crida2017}). If this were the case, they would follow the same angular speed as the perturbing object, that is, the putative planet. However, we showed (along with Ligi et al.) that those blobs follow a Keplerian motion appropriate for their separation from the star.

Finally, we note that \cite{Ligi2017} proposed that blobs B and C could be vortices (\cite{Meheut2012}). This explanation might very well be true. Another scenario might be suggested by the possibility that they are in 1:2 resonance with a putative planet related to blob D. It concerns planetesimals or asteroid giant impacts that generate dust clouds. This might be a manifestation of the general phenomenon of planetesimal erosion that is expected to follow the formation of giant planets (see, e.g., \cite{Turrini2012, Turrini2018}). However, the probability of observing such clouds is low in a gas-rich disk such as that of HD~169142 because large planetesimals are required to generate clouds as large as blobs B and C, unless the impact occurs far from the disk plane. The debris cloud from an impact roughly expands until the debris sweeps a gas mass that is no more than an order of magnitude higher than the mass of the debris itself. Because the volume of the clouds is at least one hundredth of the total volume of ring 1, this requires that the mass of the interacting bodies is higher than 1/1000 of the mass of the disk when we assume a disk gas-to-dust ratio of unity and that the impact occurs close to the disk plane. The impacting bodies should then have a mass on the order of that of Mars or at least the Moon. Since it is not likely that many such objects are present in the disk of HD~169142, the probability of observing one or even more similar debris clouds is likely very low.



\begin{acknowledgements}
The authors thank A. Pohl for allowing them to use the original reduction of the $Q_\Phi$ data set and the ESO Paranal Staff for support for conducting the observations.  E.S., R.G., D.M., S.D. and R.U.C. acknowledge support from the "Progetti Premiali" funding scheme of the Italian Ministry of Education, University, and Research. E.R. and R.L. are supported by the European Union’s Horizon 2020 research and innovation programme under the Marie Skłodowska-Curie grant agreement No 664931. This work has been supported by the project PRININAF 2016 The Cradle of Life - GENESIS-SKA (General Conditions in Early Planetary Systems for the rise of life with SKA). The authors acknowledge financial support from the Programme National de Plan\'etologie (PNP) and the Programme National de Physique Stellaire (PNPS) of CNRS-INSU. This work has also been supported by a grant from the French Labex OSUG\@2020 (Investissements d’avenir - ANR10 LABX56). The project is supported by CNRS, by the Agence Nationale de la Recherche (ANR-14-CE33-0018). This work is partly based on data products produced at the SPHERE Data Centre hosted at OSUG/IPAG, Grenoble. We thank P. Delorme and E. Lagadec (SPHERE Data Centre) for their efficient help during the data reduction process. SPHERE is an instrument designed and built by a consortium consisting of IPAG (Grenoble, France), MPIA (Heidelberg, Germany), LAM (Marseille, France), LESIA (Paris, France), Laboratoire Lagrange (Nice, France), INAF Osservatorio Astronomico di Padova (Italy), Observatoire de Genève (Switzerland), ETH Zurich (Switzerland), NOVA (Netherlands), ONERA (France) and ASTRON (Netherlands) in collaboration with ESO. SPHERE was funded by ESO, with additional contributions from CNRS (France), MPIA (Germany), INAF (Italy), FINES (Switzerland) and NOVA (Netherlands). SPHERE also received funding from the European Commission Sixth and Seventh Framework Programmes as part of the Optical Infrared Coordination Network for Astronomy (OPTICON) under grant number RII3-Ct-2004-001566 for FP6 (2004-2008), grant number 226604 for FP7 (2009-2012), and grant number 312430 for FP7 (2013-2016).
\end{acknowledgements}

%
%

\bibliographystyle{aa}

\begin{thebibliography}{73}

\bibitem[Allard et al. 2001]{Allard2001} Allard, F.,Hauschildt, P.~H.,Alexander, D.~R. et al. 2001, ApJ,556, 357
\bibitem[ALMA 2015]{ALMA2015}ALMA Partnership, Brogan, C.~L., P\'erez, et al. 2015, ApJL, 808, L3,
\bibitem[Avenhaus et al. 2018]{Avenhaus2018}Avenhaus, H., Quanz, S.~P., Garufi, et al. 2018, ApJ, 863, 44
\bibitem[Bae et al. 2018]{Bae2018}Bae, J.,Pinilla, P., Birnstiel, T., 2018, ApJL, 864, L26,
\bibitem[Bertrang et al. 2018]{Bertrang2018}Bertrang, G.~H.-M., Avenhaus, H., Casassus, S., et al. 2018, MNRAS, 474, 5105
\bibitem[Beuzit et al. 2008]{Beuzit2008}Beuzit, J.-L., Feldt, M., Dohlen, et al. 2008, SPIE, 7014, 701418
\bibitem[Biller et al. 2014]{Biller2014}Biller, B.~A., Males, J., Rodigas, et al.T.,Morzinski, K., 2014, ApJL, 792, L22
\bibitem[Blondel et al. 2006]{Blondel2006}Blondel, P.~F.~C.,Djie, H.~R.~E.~T.~A. 2006, A\&A, 456, 1045
\bibitem[Boccaletti et al. 2008]{Boccaletti2008}Boccaletti, A.,Abe, L.,Baudrand, J., Daban, J.-B., 2008, SPIE, 7015, 70151B,
\bibitem[Cardelli et al. 1989]{Cardelli1989}Cardelli, J.~A., Clayton, G.~C., Mathis, J.~S., 1989, ApJ, 345, 245
\bibitem[Chauvin et al. 2017]{Chauvin2017}Chauvin, G., Desidera, S., Lagrange, A.-M., et al. 2017, in Proceedings of the Annual meeting of the French Society of Astronomy \& Astrophysics, ed. Reyl\'e, C., Di Matteo, P., Herpin, F.,Lagadec, E., Lan\c con, A., Meliani, Z., Royer, F., p. 331 
\bibitem[Chen et al. 2018]{Chen2018}Chen, L.,K\'osp\'al, \'A.,\'Abrah\'am, P., et al. 2018, A\&A, 609, A45,
\bibitem[Claudi et al. 2008]{Claudi2008}Claudi, R.~U.,Turatto, M.,Gratton, R.~G., et al. 2008, SPIE, 7014, 70143E
\bibitem[Crida et al. 2015]{Crida2017}Crida, A., Bitsch, B., Ndugu, N., Morbidelli, A., 2017, EPSC, 11, 44
\bibitem[Currie et al. 2015]{Currie2015}Currie, T., Cloutier, R.,Brittain, S., et al. 2015, ApJL, 814, L27,
\bibitem[Currie et al. 2014]{Currie2014}Currie, T., Muto, T., Kudo, T., et al. 2014, ApJL, 796, L30,
\bibitem[Delorme et al. 2017]{Delorme2017}Delorme, P., Meunier, N., Albert, D., et al. 2017 in Proceedings of the Annual meeting of the French Society of Astronomy \& Astrophysics,, ed. Reyl\'e, C., Di Matteo, P., Herpin, 
F.,Lagadec, E.,	Lan\c con, A., Meliani, Z., Royer, F., p. 347
\bibitem[Dent et al. 2013]{Dent2013}Dent, W.~R.~F., Thi, W.~F., Kamp, I., et al. 2013, PASP, 125, 477
\bibitem[Dipierro et al. 2018]{Dipierro2018}Dipierro, G., Ricci, L., P\'erez, L., et al. 2018, MNRAS, 475, 5296
\bibitem[Dohlen et al. 2008]{Dohlen2008}Dohlen, K., Langlois, M., Saisse, M., et al. 2008, SPIE, 7014, 70143L,
\bibitem[Dunkin et al. 1997a]{Dunkin1997a}Dunkin, S.~K., Barlow, M.~J., Ryan, S.~G., 1997a, MNRAS, 286, 604
\bibitem[Dunkin et al. 1997b]{Dunkin1997b}Dunkin, S.~K., Barlow, M.~J., Ryan, S.~G., 1997b, MNRAS, 290, 165
\bibitem[Dong \& Fung 2017]{Dong2017}Dong, R. \& Fung, J. 2017 ApJ, 835, 146
\bibitem[Dong et al. 2018]{Dong2018}Dong, R., Najita, J.~R., Brittain, S., 2018, ApJ. 862, 103
\bibitem[Fedele et al. 2017]{Fedele2017}Fedele, D., Carney, M., Hogerheijde, et al. 2017, A\&A, 600, A72
\bibitem[Fung et al. 2015]{Fung2015}Fung, J., \& Dong, R., 2015, ApJL, 815, L21
\bibitem[Grady et al. 2007]{Grady2007}Grady, C.~A., Schneider, G., Hamaguchi, K., et al. 2007, ApJ, 665, 1391
\bibitem[Honda et al. 2012]{Honda2012}Honda, M., Maaskant, K., Okamoto, Y.~K., et al. 2012, ApJ, 752, 143
\bibitem[Kanagawa et al. 2016]{Kanagawa2016}Kanagawa, K.~D., Muto, T., Tanaka, H., et al. 2016, PASJ, 68, 43
\bibitem[Keppler et al. 2018]{Keppler2018}Keppler, M., Benisty, M., M\"uller, et al. 2018, A\&A, 617, A44
\bibitem[Klacka et al. 1993]{Klacka1993}Klacka, J., \& Saniga, M., 1993, EM\&P, 60, 23
\bibitem[Kraus et al. 2012]{Kraus2012}Kraus, A.~L., \& Ireland, M.~J., 2012, ApJ, 745, 5
\bibitem[Lazareff et al. 2017]{Lazareff2017}Lazareff, B., Berger, J.-P., Kluska, J., et al., 2017, A\&A, 588, 85
\bibitem[Ligi et al. 2017]{Ligi2017}Ligi, R., Vigan, A., Gratton, R., et al. 2018, MNRAS, 473, 1774
\bibitem[Maaskan et al. 2013]{Maaskant2013}Maaskant, K.~M., Honda, M., Waters, L.~B.~F.~M., et al. 2013, A\&A, 555, A64
\bibitem[Machida et al. 2010]{Machida2010}Machida, M.~N., Kokubo, E.,I nutsuka, S.-I., Matsumoto, T., 2010, MNRAS, 405, 1227
\bibitem[Macias et al. 2017]{Macias2017}Macias, E., Anglada, G., Osorio, M., et al., 2017, ApJ, 838, 97
\bibitem[Macintosh et al. 2014]{Macintosh2014}Macintosh, B., Graham, J.~R., Ingraham, P., et al. 2014, Proceedings of the National Academy of Science, 111, 12661
\bibitem[Manoj et al. 2005]{Manoj2005}Manoj, P., \& Bhatt, H.~C., 2005, A\&A, 429, 525
\bibitem[Manoj et al. 2007]{Manoj2007}Manoj, P., Ho, P.~T.~P., Ohashi, N., et al. 2007, ApJL, 667, L187
\bibitem[Mawet et al. 2014]{Mawet2014}Mawet, D., Milli, J., Wahhaj, Z., et al. 2014, ApJ, 792, 97
\bibitem[Meheut et al. 2012]{Meheut2012}Meheut, H., Meliani, Z., Varniere, P., Benz, W., 2012,A\&A, 545, A134
\bibitem[Mesa et al. 2015]{Mesa2015}Mesa, D., Gratton, R., Zurlo, A., et al. 2015, A\&A, 576, A121
\bibitem[Monnier et al. 2017]{Monnier2017}Monnier, J.~D., Harries, T.~J., Aarnio, A., 2017, ApJ, 838, 20
\bibitem[Muller et al. 2018]{Muller2018}M\"uller, A., Keppler, M., Henning, T., et al. 2018, A\&A, 617, L2
\bibitem[Murphy et al. 2015]{Murphy2015}Murphy, S.~J., Corbally, C.~J., Gray, R.~O., et al. 2015, PASA, 32, 36
\bibitem[Nowak et al. 2018]{Nowak2018}Nowak, M., Le Coroller, H., Arnold, L., et al. 2018,  A\&A, 615, A144
\bibitem[Osorio et al. 2014]{Osorio2014}Osorio, M., Anglada, G., Carrasco-Gonz\'alez, C., et al. 2014, ApJ, 791, L360
\bibitem[Panic et al. 2010]{Panic2010}Pani\'c, O., van Dishoeck, E.~F., Hogerheijde, M.~R., et al. 2010, A\&A, 519, A110
\bibitem[Paunzen et al. 2001]{Paunzen2001}Paunzen, E., Duffee, B., Heiter, U., Kuschnig, R., Weiss, W.~W., 2001, A\&A, 373, 625
\bibitem[Pavlov et al. 2008]{Pavlov2008}Pavlov, A., M\"oller-Nilsson, O., Feldt, M., et al. 2008, SPIE, 7019, 701939
\bibitem[Pecaut et al. 2013]{Pecaut2013}Pecaut, M.~J., \& Mamajek, E.~E., 2013, ApJS, 208, 9
\bibitem[Pinte et al. 2018]{Pinte2018}Pinte, C., Price, D.~J., M\'enard, F., et al. 2018, ApJ, 860, L13
\bibitem[Pohl et al. 2017]{Pohl2017}Pohl, A., Benisty, M., Pinilla, P., et al. 2017, ApJ, 850, 52
\bibitem[Quanz et al. 2013a]{Quanz2013a}Quanz, S.~P., Amara, A., Meyer, M.~R., et al. 2013a, ApJL, 766, L1
\bibitem[Quanz et al. 2013b]{Quanz2013b}Quanz, S.~P., Avenhaus, H., Buenzli, E., et al. 2013b, ApJL, 766, L2
\bibitem[Quanz et al. 2015]{Quanz2015}Quanz, S.~P., Amara, A., Meyer, M.~R., et al. 2015, ApJ, 807, 64
\bibitem[Raman et al. 2006]{Raman2006}Raman, A., Lisanti, M., Wilner, D.~J., Qi, C., Wogerheijde, M., 2006, AJ, 131, 2290
\bibitem[Rameau et al. 2017]{Rameau2017}Rameau, J., Follette, K.~B., Pueyo, L., et al. 2017, AJ, 153, 244
\bibitem[Reggiani et al. 2014]{Reggiani2014}Reggiani, M., Quanz, S.~P., Meyer, M.~R., et al. 2014, ApJL, 792, L23
\bibitem[Reggiani et al. 2018]{Reggiani2018}Reggiani, M., Christiaens, V., Absil, O., et al. 2018, A\&A, 611, A74
\bibitem[Salyk et al. 2013]{Salyk2013}Salyk, C., Herczeg, G.~J., Brown, J.~M., et al. 2013, ApJ, 769, 21
\bibitem[Sallum et al. 2015]{Sallum2015}Sallum, S., Follette, K.~B., Eisner, J.~A., et al. 2015, Nature, 527, 342
\bibitem[Sissa et al. 2018]{Sissa2018}Sissa, E., Gratton, R., Garufi, A., et al. 2018, A\&A, 619, A160
\bibitem[Soummer et al. 2012]{Soummer2012}Soummer, R., Pueyo, L., Larkin, J., 2012, ApJL, 755, L28
\bibitem[Teague et al. 2018]{Teague2018}Teague, R., Bae, J., Bergin, E.~A., Birnstiel, T., Foreman-Mackey, D., 2018, ApJL, 860, L12
\bibitem[Turrini et al. 2012]{Turrini2012}Turrini, D., Coradini, A., Magni, G., 2012, ApJ, 750, 8
\bibitem[Turrini et al. 2018]{Turrini2018}Turrini, D., Marzari, F., Polychroni, D., Testi, L., 2018, arXiv1802.04361,
\bibitem[Vieira et al. 2003]{Vieira2003}Vieira, S.~L.~A., Corradi, W.~J.~B., Alencar, S.~H.~P., et al. 2003, AJ, 126, 2971
\bibitem[Vigan et al. 2010]{Vigan2010}Vigan, A., Moutou, C., Langlois, M., et al. 2010, MNRAS, 407, 71
\bibitem[Vioque et al. 2010]{Vioque2018}Vioque, M., Oudmaijer, R.~D., Baines, D., et al. 2018, A\&A, 620, A128
\bibitem[Wagner et al. 2015]{Wagner2015}Wagner, K.~R., Sitko, M.~L., Grady, C.~A., et al. 2005, ApJ, 798, 94
\bibitem[Wagner et al. 2018]{Wagner2018}Wagner, K., Follete, K.~B., Close, L.~M., et al. 2018, ApJ, 863, L8
\bibitem[Wright 2003]{Wright2003}Wright, A., 2003, Nature, 422, 130
\bibitem[Zhu et al. 2015]{Zhu2015}Zhu, Z., Dong, R., Stone, J.~M., Rafikov, R.~R. 2015, ApJ, 813, 88

\end{thebibliography}

\section{Appendix: Radial and rotational velocities for HD~169142}

We measured relative radial velocities and projected rotational velocities of HD~169142 from a series of eight high-resolution archive spectra of HD~169142 acquired with the HARPS spectrograph at the ESO 3.6 m telescope in La Silla in 2008 (Program 080.C-0712, PI: Desort). The spectra were reduced using the ESO pipeline. Radial velocities were obtained by cross-correlating them with the average of the last two spectra that have the highest S/N. The radial velocities are then relative. Relevant data are given in Table~\ref{t:vrad}.

\begin{table}
\caption{Radial velocities and $V~\sin{i}$}
\begin{tabular}{lccc}
\hline
\hline
JD & $V_{\rm rad}$ & $V~\sin{i}$ \\
   & (km/s) & (km/s) \\
\hline
54542.286 & $-0.89\pm 0.11$ & 47.54 \\
54542.296 & $-0.97\pm 0.10$ & 48.40 \\
54542.365 & $-1.41\pm 0.08$ & 53.56 \\
54542.375 & $-1.50\pm 0.08$ & 54.18 \\
54546.281 & $-0.21\pm 0.09$ & 49.04 \\
54546.292 & $-0.20\pm 0.09$ & 49.33 \\
54546.361 & $-0.09\pm 0.08$ & 50.24 \\
54546.372 & $~0.09\pm 0.08$ & 50.22 \\
\hline
\end{tabular}
\label{t:vrad}
\end{table}

\end{document}